\documentclass[twocolumn,prl,superscriptaddress,preprintnumbers,amsmath,amssymb,showpacs]{revtex4}

\usepackage{graphicx}
\usepackage{dcolumn}
\usepackage{bm}
\begin{document}
\title{Theory of Resonant Inelastic X-ray Scattering by Collective Magnetic Excitations}

\author{M. W. Haverkort}
  \affiliation{Max Planck Institute for Solid State Research, Heisenbergstra{\ss}e 1, D-70569 Stuttgart Germany}

\date{\today}

\begin{abstract}
I present a tractable theory for the Resonant Inelastic X-ray Scattering (RIXS) of magnons. The low-energy transition operator is written as a product of local spin operators and fundamental x-ray absorption spectral functions. This leads to simple selection rules. The scattering cross section linear (quadratic) in spin operators is proportional to the fundamental magnetic circular (linear) dichroic spectral function. RIXS is a novel tool to measure magnetic quasi particles (magnons) and the incoherent spectral weight, as well as multiple magnons up to very high energy losses, in small samples, thin films and multilayers, complementary to Neutron scattering.

\end{abstract}


\maketitle

Magnetic excitations in solids are traditionally investigated with inelastic neutron scattering. This technique has led to a better understanding of magnetism in general and shaped the thinking of physicists in terms of magnons as the magnetic quasi-particles present within a solid. It is important that the interaction of neutrons with the magnetic moment is well understood and can be approximated by a function linear in spin operators.

One might expect that x-rays are not suited for the study of magnetic excitations, as photons, contrary to neutrons, do not carry a magnetic spin and therefore cannot directly excite single magnetic excitations. Recent resonant inelastic x-ray scattering (RIXS) experiments see clear dispersing magnetic excitations \cite{Hill08,Bracovich09,Schlappa09,Ghiringhelli10}. More high quality data is expected due to the experimental improvement in both brilliance and resolution \cite{Huotari06, Ghiringhelli06}. For neutron scattering the magnetic dipole interaction between the neutron and electron spin is well understood. For RIXS the effective magnetic interaction is highly non-trivial. The optical dipole transition does not introduce spin-flip excitations. The magnetic excitations are a result of core-hole spin-orbit coupling and many-body interactions active in the resonant intermediate state \cite{Groot98,Luo93,Veenendaal06,Ament09,GrootKotani08,Chubukov95,Devereaux03,Brink06}.

Based on local cluster calculations De Groot \textit{et al.} \cite{Groot98} realized that magnetic excitations are in principle possible. Luo \textit{et al.} \cite{Luo93,Veenendaal06} showed that, within the fast collision approximation, it is possible to describe the RIXS spectra in terms of a low-energy scattering operator, which can excite single magnons. At the Cu $L_{2,3}$ edge this approximation allows one to describe the magnetic excitations including their momentum and polarization dependence \cite{Veenendaal06,Ament09}. In general however the RIXS spectra show not only a strong momentum and polarization dependence, but also a strong resonant energy dependence \cite{Hill08,GrootKotani08}. In order to include this, one has to take into account the interactions in the intermediate state, without neglecting the momentum dependence of the final state.

Within this letter I present an effective low-energy operator for RIXS, focused on magnetic excitations. The idea of the derivation, is not to make approximations to the Hamiltonian, or to the Green's functions used to evaluate the RIXS spectral function. Using symmetry arguments an exact solution for magnetic excitations can be obtained by factorizing the RIXS transition operator into symmetric polynomials of spin operators and fundamental x-ray absorption spectral functions. The fundamental spectral functions describe the dependence of the x-ray absorption on the magnetic moment $\boldsymbol{m}$. These are the isotropic, magnetic circular dichroic (XMCD) and magnetic linear dichroic (XMLD) spectral functions \cite{Laan93}.

I will start the derivation with the Kramers-Heisenberg formula for the double differential cross section:
\begin{eqnarray}
\label{FormulaKramersHeisenberg}
 \nonumber\frac{\delta^2\sigma}{\delta\Omega\delta\omega} &\propto& \lim_{\Gamma \to 0^+} \sum_f\left|\langle f|T^{\dag}_{\varepsilon_{o}}\frac{1}{\omega_{i}+E_i+\imath\Gamma/2-H}T_{\varepsilon_{i}}|i\rangle\right|^2\\
&&\qquad\delta(\omega_{i}-\omega_{o}+E_i-E_f),
\end{eqnarray}
where $i$ ($f$) labels the set of initial (final) states, $E_i$ is the energy of state $|i\rangle$, $H$ is the Hamiltonian, and $\Gamma$ is the life-time broadening, which, if all states are included in the Hamiltonian goes to zero. In practice one only keeps the relevant states in the basis set of $H$. The influence of the neglected states is included by a possible state-dependent non-zero value of $\Gamma$. The photon energy and the polarization of the incoming (outgoing) photon is given by $\omega_{i(o)}$ and $\varepsilon_{i(o)}$. $T_{\epsilon}$ is the optical transition operator given by the perturbation of the photon field on the system: $T_{\epsilon}=\boldsymbol{p}\cdot\boldsymbol{A}$, with $\boldsymbol{A}$ the photon vector field and $\boldsymbol{p}$ the electron momentum operator. The state $|i\rangle$ represents the ground-state, for example a magnetically ordered N{\'e}el state. The operator $T$ creates a core-hole and adds a conduction electron to the system. The state given by $T |i\rangle$ is in general not an eigen-state of H and it will fluctuate in time. The Green's function $(\omega_i+E_i+\imath \Gamma/2 -H)^{-1}$ describes this propagation. A possible propagation path that would add a magnetic excitation to the system could be the flip of the core spin due to the core-hole spin-orbit coupling. Many other excitations are possible. After the Green's function has acted on the state $T|i\rangle$ the operator $T^\dag$ fills the core hole and at the same time annihilates a valence electron. This entire process might leave the system in an excited state.

Using $\delta(\omega+E_i-E_f)\propto \Im [ \lim_{\Gamma \to 0^+} (\omega+E_i-E_f+\imath \Gamma/2)^{-1} ]$ one can introduce the following notation:
\begin{eqnarray}
\label{FormulaSusceptibility}
\frac{\delta^2\sigma}{\delta\Omega\delta\omega} &\propto& -\Im [\chi(\omega)]\\
\nonumber \chi(\omega) &=& \lim_{\Gamma \to 0^+} \langle i | {R^{\varepsilon_{i}\varepsilon_{o}}_{\omega_i}}^{\dag} \frac{1}{\omega+E_i+\imath \Gamma/2 - H} R^{\varepsilon_{i}\varepsilon_{o}}_{\omega_i} |i\rangle,
\end{eqnarray}
with $\omega=\omega_{i}-\omega_{o}$, and:
\begin{equation}
 R^{\varepsilon_{i}\varepsilon_{o}}_{\omega_i}=T_{\varepsilon_{o}}^{\dag} \frac{1}{\omega_i+E_i+\imath\Gamma/2-H} T_{\varepsilon_{i}}.
\end{equation}
The operator $R^{\varepsilon_{i}\varepsilon_{o}}_{\omega_i}$ is the low-energy effective scattering operator. The operator both creates and annihilates a core hole, leaving the system in a low-energy excited state. The $\omega_i$ dependence of $R^{\varepsilon_{i}\varepsilon_{o}}_{\omega_i}$ describes the true resonant nature of RIXS. The Green's function $(\omega+E_i+\imath\Gamma/2-H)^{-1}$ in Eq. \ref{FormulaSusceptibility} describes how these low-energy excitations, for example a spin-flip, propagate through the sample.

It is useful to factorize the momentum dependence of $R^{\varepsilon_{i}\varepsilon_{o}}_{\omega_i}$. The photon field, $\boldsymbol{A}$, can be expanded around each atom at site $\boldsymbol{r}_j$. One should take into account that $\boldsymbol{A}$ changes phase when one changes lattice position. For photon fields described by a plane wave the phase shift is given by a simple exponential:
\begin{eqnarray}
T_{\varepsilon}=\boldsymbol{p} \cdot \boldsymbol{A} = \sum_j e^{\imath \boldsymbol{k} \cdot \boldsymbol{r}_j} \boldsymbol{p}_j \cdot \boldsymbol{A}_j = \sum_j e^{\imath \boldsymbol{k} \cdot \boldsymbol{r}_j} T_{j,\varepsilon},
\end{eqnarray}
where the sum is over all atoms, $\boldsymbol{k}$ is the photon wave vector, $\boldsymbol{r}_{j}$ is the position of atom $j$ and $\boldsymbol{p}_j$ ($\boldsymbol{A}_j$) is the momentum operator (photon vector field) expanded around atom $j$.

$T_{j,\varepsilon}$ ($T_{j',\varepsilon}^{\dag}$) creates (annihilates) a core hole at site $j$ ($j'$). For core level spectroscopy one can assume that the core hole does not hop from one site to another. The site where the core hole is created is thus the same as the site where the core hole is annihilated. Inserting this in the definition of $R^{\varepsilon_{i}\varepsilon_{o}}_{\omega_i}$ yields:
\begin{eqnarray}
\label{eqRqj}
\nonumber R^{\varepsilon_{i}\varepsilon_{o}}_{\omega_i}&=&\sum_{j',j} e^{\imath (\boldsymbol{k}_i \cdot \boldsymbol{r}_j - \boldsymbol{k}_o \cdot \boldsymbol{r}_{j'})} T_{j',\varepsilon_{o}}^{\dag} \frac{1}{\omega_{i}+E_i+\imath\Gamma/2-H} T_{j,\varepsilon_{i}}\\
\nonumber &=&  \sum_{j} e^{\imath (\boldsymbol{q} \cdot \boldsymbol{r}_{j})} T_{j,\varepsilon_{o}}^{\dag} \frac{1}{\omega_{i}+E_i+\imath\Gamma/2-H} T_{j,\varepsilon_{i}}\\
 &=&  \sum_{j} e^{\imath (\boldsymbol{q} \cdot \boldsymbol{r}_{j})} R_{\omega_i,j}^{\varepsilon_i\varepsilon_o} = R^{\varepsilon_{i}\varepsilon_{o}}_{\omega_i,\boldsymbol{q}},
\end{eqnarray}
with $\boldsymbol{q}=\boldsymbol{k}_i-\boldsymbol{k}_o$. 

Before factorizing the effective scattering operator $R^{\varepsilon_{i}\varepsilon_{o}}_{\omega_i,\boldsymbol{q}}$ into spin operators and fundamental x-ray absorption spectral functions I will address a few general properties of $R^{\varepsilon_{i}\varepsilon_{o}}_{\omega_i,\boldsymbol{q}}$. It is interesting to note that the conductivity tensor at x-ray energies is defined such that: $\epsilon_o^* \cdot\sigma\cdot\epsilon_i = \langle i|R^{\varepsilon_{i}\varepsilon_{o}}_{\omega_i,\boldsymbol{q}}|i\rangle$. The absorption for a given polarization is proportional to $-\Im[\epsilon^* \cdot\sigma\cdot\epsilon]$ at $\boldsymbol{q}=0$. In order to calculate the inelastic scattering between state $|i\rangle$ and state $|f\rangle$ one needs to evaluate $\langle f|R^{\varepsilon_{i}\varepsilon_{o}}_{\omega_i,\boldsymbol{q}}|i\rangle$. Using that $ 2 \langle f | R^{\varepsilon_{i}\varepsilon_{o}}_{\omega_i,\boldsymbol{q}} |i\rangle = \langle f+i |R^{\varepsilon_{i}\varepsilon_{o}}_{\omega_i,\boldsymbol{q}}| f+i \rangle + \imath \langle \imath f+i |R^{\varepsilon_{i}\varepsilon_{o}}_{\omega_i,\boldsymbol{q}}| \imath f+i \rangle - (1+\imath)(\langle i |R^{\varepsilon_{i}\varepsilon_{o}}_{\omega_i,\boldsymbol{q}}| i \rangle + \langle f |R^{\varepsilon_{i}\varepsilon_{o}}_{\omega_i,\boldsymbol{q}}| f \rangle) $ one realizes that the RIXS transition probability for making excitations from state $|i\rangle$ to state $|f\rangle$ is determined by x-ray absorption spectral functions of the initial state, the final state and linear combinations of these two states. For magnetic excitations $|i\rangle$ could be a state with spin down, $|f\rangle$ a state with spin up. The real and complex linear combinations are states with rotated spin directions. The fundamental spectral functions describe the dependence of the x-ray absorption on the magnetization direction of a given system. For magnetic excitations the RIXS transition probability between state $|i\rangle$ and state $|f\rangle$ is thus given by some linear combination of fundamental spectral functions. Below I will derive this dependence exactly with the use of symmetry arguments.

For magnetic excitations one can restrict the basis set of $R^{\varepsilon_{i}\varepsilon_{o}}_{\omega_i,\boldsymbol{q}}$ to the ground-state and all states with all possible spin excitations from the ground-state. Within this basis one can use the theorem of operator equivalence as introduced by Stevens \cite{Stevens52} for the description of crystal-fields in Rare-Earth compounds. Any operator acting in spin-space only can be written as a polynomial of spin operators. $R^{\varepsilon_{i}\varepsilon_{o}}_{\omega_i,j}=P(S)$, where $P$ remains to be determined.

The terms in $P$ are restricted by symmetry relations. I will first assume spherical symmetry and later branch down to the possible crystal symmetries. In general $R^{\varepsilon_{i}\varepsilon_{o}}_{\omega_i,j}$ must be a scalar. In spherical symmetry $R^{\varepsilon_{i}\varepsilon_{o}}_{\omega_i,j}$ can only depend on the angle between $\epsilon_i$ and $\epsilon_o^*$. Operators of the form $\epsilon_0^* \cdot \epsilon_i$, $\epsilon_0^* \times \epsilon_i \cdot S$, $(\epsilon_0^* \cdot S) (\epsilon_i \cdot S) + (\epsilon_i \cdot S) (\epsilon_0^* \cdot S)$ and $(\epsilon_0^* \cdot \epsilon_i) (S \cdot S)$ are allowed.

The pre-factors of these different symmetry-allowed terms can be obtained using the notion that the RIXS spectral function for zero energy loss should reproduce the elastic scattering. The equation for the elastic scattering follows from the Kramers-Heisenberg equation by restricting the possible excitations to the ground-state:
\begin{eqnarray}
\nonumber && \chi(\boldsymbol{q},\omega\to0)=\\
\nonumber && \lim_{\Gamma \to 0^+}\langle i|{R^{\varepsilon_{i}\varepsilon_{o}}_{\omega_i,\boldsymbol{q}}}^{\dag}|i\rangle\langle i| \frac{1}{\omega+E_i+\imath\Gamma/2-H}|i\rangle\langle i| R^{\varepsilon_{i}\varepsilon_{o}}_{\omega_i,\boldsymbol{q}}|i\rangle\Rightarrow\\
&& \frac{\delta\sigma}{\delta\Omega} \propto \left|\langle i| R^{\varepsilon_{i}\varepsilon_{o}}_{\omega_i,\boldsymbol{q}}|i\rangle\right|^2.
\label{eqElastic}
\end{eqnarray}

At the same time, for elastic scattering one can take advantage of the optical theorem which states that the elastic scattering is related to the x-ray absorption spectral function \cite{Newton76}. The elastically scattered signal is proportional to $|\epsilon_{o}^* \cdot\sigma\cdot\epsilon_{i}|^2$, confirming the relation between $ R^{\varepsilon_{i}\varepsilon_{o}}_{\omega_i,\boldsymbol{q}}$ and the conductivity tensor. Hannon \textit{et al.} \cite{Hannon88} related the elastic intensity to the magnetization direction; this can be used to determine the pre-factors for the inelastic scattering. The elastic scattering is given by \cite{Hannon88,Haverkort10}:
\begin{align}
&\langle i| R^{\varepsilon_{i}\varepsilon_{o}}_{\omega_i,j} |i\rangle=\sigma^{(0)} \varepsilon_i \cdot \varepsilon_o^* + \sigma^{(1)} \varepsilon_i \times \varepsilon_o^* \cdot \boldsymbol{m}_j \\
\nonumber&                         + \sigma^{(2)} ((\varepsilon_i \cdot \boldsymbol{m}_j) (\varepsilon_o^* \cdot \boldsymbol{m}_j)-\frac{1}{3}\varepsilon_i\cdot\varepsilon_o^*),
\end{align}
where $\sigma^{(0)}$ is the numerical value of the complex isotropic spectral function, $\sigma^{(1)}$ and $\sigma^{(2)}$ are the complex, fundamental, XMCD and XMLD spectral functions respectively and $\boldsymbol{m}_j$ is a unit vector in the direction of magnetization. The operator $R^{\varepsilon_{i}\varepsilon_{o}}_{\omega_i,j}$ is given in terms of symmetric spin operators as:
\begin{align}
\label{Rjee}
&R^{\varepsilon_{i}\varepsilon_{o}}_{\omega_i,j}=\sigma^{(0)} \varepsilon_i \cdot \varepsilon_o^* + \frac{\sigma^{(1)}}{s} \varepsilon_o^* \times \varepsilon_i \cdot S_j \\
\nonumber&                         + \frac{\sigma^{(2)}}{s(2s-1)} (\varepsilon_i \cdot S_j\ \varepsilon_o^* \cdot S_j+\varepsilon_o^* \cdot S_j\ \varepsilon_i \cdot S_j-\frac{2}{3}\varepsilon_i\cdot\varepsilon_o^*S_j^2),
\end{align}
where $S_j$ is the spin operator acting at site $j$ and $s$ the expectation value $\langle S_j^2 \rangle=s(s+1)$. For real crystals the local point-group has to be included. This leads to a branching of the fundamental spectral functions \cite{Laan93} and polynomials of higher orders in S, up to an order of 2s \cite{Haverkort10}. Eq. \ref{Rjee} is the main result of the present letter and an exact form of the RIXS transition operator for magnetic excitations truncated to single site transitions. It leads to beautifully simple selection rules: One can measure single spin-flip transitions with cross polarized light for spins in the plane of the polarizations. The operator quadratic in $S$ can lead to single ($\Delta S=1$) and double  ($\Delta S=2$) spin-flip transitions, as well as contributions to the elastic line. For $S=1/2$ systems $\sigma^{(2)}$ is zero by symmetry.
 
In order to clarify the method I present RIXS spectra at the $L_{2,3}$ edge of samples containing Cu$^{2+}$ in $D_{4h}$ symmetry and Ni$^{2+}$ in $O_{h}$ symmetry. I will use linearized spin-wave theory for a Heisenberg-model on a 1D chain, a 2D square and a 3D cubic lattice with nearest neighbor interactions. Linear spin-wave theory assumes local ordered moments. For perfect one dimensional samples the ground-state is not ordered and for $S=1/2$ systems the magnetic excitations are two spinon excitations \cite{Majlis00}. It is often more transparent to write $R^{\varepsilon_{i}\varepsilon_{o}}_{\omega_i,j}$ as the inner-product of two polarization vectors and a $3\times3$ tensor. In such a matrix notation, $R^{\varepsilon_{i}\varepsilon_{o}}_{\omega_i,j}$ for Cu$^{2+}$, a $S=1/2$ system, in tetragonal symmetry becomes:
\begin{equation}
\label{TRShalf}
R^{\varepsilon_{i}\varepsilon_{o}}_{\omega_i,j}=
\bm{\varepsilon}_{o}^*\cdot\left(
 \begin{array}{ccc}
   \sigma^{(0)}_{a_{1g}^B}   & 2 S_z \sigma^{(1)}_{a_{2u}} & -2 S_y \sigma^{(1)}_{e_u} \\
  -2 S_z \sigma^{(1)}_{a_{2u}} & \sigma^{(0)}_{a_{1g}^B}   &  2 S_x \sigma^{(1)}_{e_u} \\
   2 S_y \sigma^{(1)}_{e_u}  & -2 S_x \sigma^{(1)}_{e_u} & \sigma^{(0)}_{a_{1g}^A} \\
 \end{array}
\right)\cdot\bm{\varepsilon}_{i}.
\end{equation}

The fundamental spectra of Cu$^{2+}$ determining the RIXS transition at the $L_{2,3}$ edge can be seen in Fig. 1a. They show a striking feature; the $\sigma^{(0)}_{a_{1g}^A}$ and $\sigma^{(1)}_{e_u}$ spectra are zero. This is a direct consequence of having only one hole with $x^2$-$y^2$ symmetry in which one can not excite a $p$ electron with $z$ polarized light. With RIXS one can not measure spin-flip excitations for fully aligned spins with $S_z=\pm1/2$ oriented perpendicularly to the $d_{x^2-y^2}$ orbital \cite{Veenendaal06}. As shown recently, direct spin-flip scattering is allowed in all other situations \cite{Ament09}. For real systems magnetic excitations will always be measurable as $S=1/2$ fully aligned spins never exist in a solid.

 \begin{figure*}
    \includegraphics[width=1.0\textwidth]{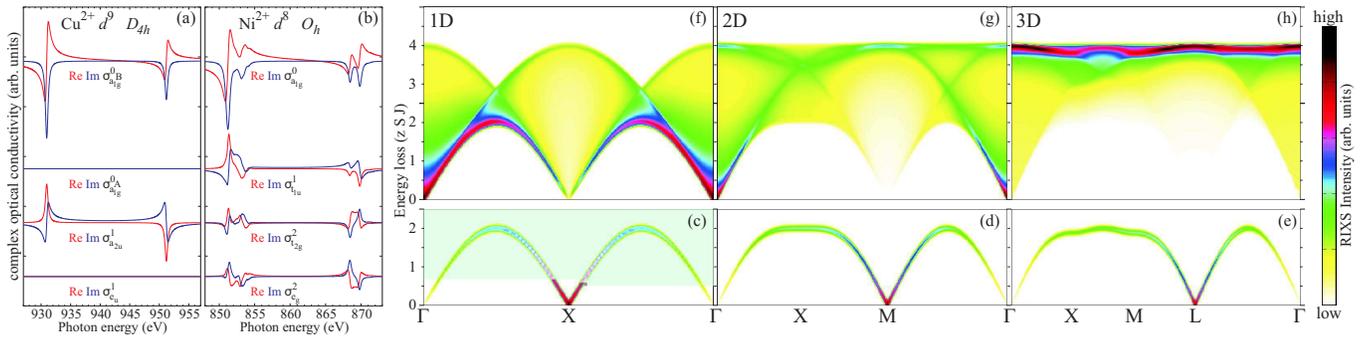}
    \caption{(color online) Left: Fundamental x-ray absorption spectra that enter into the RIXS transition operator as energy dependent complex matrix elements calculated for (a) Cu$^{2+}$ and (b) Ni$^{2+}$. Right: The Cu$^{2+}$ and Ni$^{2+}$ one magnon (c,d,e) and Ni$^{2+}$ two magnon (f,g,h) RIXS spectral function, calculated using linear spin-wave theory for a 1D chain (c,f), a 2D square (d,g) and a 3D cubic (e,h) Heisenberg model in energy loss units of $z\,S\,J$ (number of neighbors $\times$ spin $\times$ exchange constant).}
    \label{FigureSigma}
    \label{FigureSpinWaves}
  \end{figure*}

Knowing the transition probability one can calculate the RIXS spectral function, using for example linearized spin-wave theory \cite{Majlis00}. In Fig. 1c-e I show the spectral function as one would measure with RIXS in the cuprates. One finds that the intensity goes to zero at $\Gamma$ and diverges at the antiferromagnetic Bragg peak. This is a well known behaviour for collective excitations \cite{Majlis00}. The calculations for a two dimensional square lattice are in good agreement with resent measurements on La$_{2}$CuO$_{4}$ \cite{Ghiringhelli10}.

For Ni$^{2+}$ in cubic symmetry (a $S=1$ system) the resonant inelastic scattering transition matrix becomes:
\begin{widetext}
\begin{equation}
R^{\varepsilon_{i}\varepsilon_{o}}_{\omega_i,j}=
\bm{\varepsilon}_{o}^*\cdot\left(
 \begin{array}{ccc}
   \sigma^{(0)}_{a_{1g}} + 2(S_x^2-\frac{1}{3} S^2) \sigma^{(2)}_{e_g} & S_z \sigma^{(1)}_{t_{1u}} + (S_xS_y+S_yS_x) \sigma^{(2)}_{t_{2g}} & -S_y \sigma^{(1)}_{t_{1u}} + (S_zS_x+S_xS_z) \sigma^{(2)}_{t_{2g}} \\
  -S_z \sigma^{(1)}_{t_{1u}} + (S_xS_y+S_yS_x) \sigma^{(2)}_{t_{2g}} & \sigma^{(0)}_{a_{1g}} + 2(S_y^2-\frac{1}{3} S^2) \sigma^{(2)}_{e_g} &  S_x \sigma^{(1)}_{t_{1u}} + (S_yS_z+S_zS_y) \sigma^{(2)}_{t_{2g}} \\
   S_y \sigma^{(1)}_{t_{1u}} + (S_zS_x+S_xS_z) \sigma^{(2)}_{t_{2g}} & -S_x \sigma^{(1)}_{t_{1u}} + (S_yS_z+S_zS_y) \sigma^{(2)}_{t_{2g}} & \sigma^{(0)}_{a_{1g}} + 2(S_z^2-\frac{1}{3} S^2) \sigma^{(2)}_{e_g} \\
 \end{array}
\right)\cdot\bm{\varepsilon}_{i}.
\end{equation}
\end{widetext}

The numerical values of $\sigma$ at the $L_{2,3}$ edge for Ni$^{2+}$ are plotted in Fig. 1b. There are four fundamental spectra as the XMLD ($\sigma^{(2)}$) spectra branch to two different representations ($\sigma^{(2)}_{e_g}$ and $\sigma^{(2)}_{t_{2g}}$). In order to separate the different contributions one can use different scattering geometries or different resonant energies. One should realize that also interference terms between the different channels have to be considered. The fundamental x-ray absorption spectra are well known and can be calculated with quite good accuracy, simplifying the disentanglement of the different scattering channels.

In Fig. 1c-h, I show the one (panels c-e) and two (panels f-h) magnon intensity as one could measure with RIXS in Ni$^{2+}$ samples. For real geometries one would measure a linear combination of the two spectral functions depending on resonance energy and polarization as discussed above. For the single magnon spectral function only a single peak is seen at every momentum. For the two magnon spectral function a band is observed with the requirement that the sum of the momenta for the two magnons is equal to the transferred $\boldsymbol{q}$. One therefore might expect not to see any $\boldsymbol{q}$ dependence. The single magnon intensity diverges at the momentum corresponding to the antiferomagnetic order. This leads to an observable dispersion for the two magnon spectral function in one dimensional samples (Fig. 1f). For a three dimensional sample, the main peak of the two magnon spectra shows hardly any detectable momentum dependence (Fig. 1h), in good agreement with recent measurements on NiO \cite{Ghiringhelli09}. The one magnon branch does disperse.

In conclusion, I have presented a tractable theory for the calculation of RIXS spectral functions. It fully incorporates the intermediate state Hamiltonian including the core-hole spin orbit coupling and the multiplet features as well as the momentum dependence of the low-energy excited states. This is achieved by factorizing the RIXS spectral function into two parts for which different approximations are used. The low-energy Green's function can be calculated within any of the standard approximations available to describe spin-waves or other excitations at hand. The transition operator is written as local operators multiplied by fundamental x-ray absorption spectra. The transition operator has a simple form that allows one to determine selection rules depending on the polarization of the incoming and outgoing light. RIXS measurements can provide us with detailed information on both the magnons and the incoherent spectral weight, which contains information on the momentum dependent interactions of magnons with themselves and other degrees of freedom in the system. The RIXS cross section at the TM $L_{2,3}$ edge is so large that the technique can be used to measure small samples, thin-films or multi-layers making it possible to measure magnetic excitations in a whole new class of materials. 

I would like to thank Giniyat Khaliullin, Giacomo Ghiringhelli and Vladimir Hinkov for valuable discussions.

\end{document}